# Electron cooling by carbon monoxide in the atmospheres of Mars and Venus

Laurence Campbell* and Michael J Brunger

Address: ARC Centre for Antimatter-Matter Studies, SoCPES, Flinders University, PO box 2100, Adelaide SA 5001, Australia

Email: Laurence Campbell* - laurence.campbell@flinders.edu.au; Michael J Brunger - michael.brunger@flinders.edu.au

* Corresponding author

## Abstract

Electron cooling, in which free electrons lose energy to vibrational excitation of gases, has been identified as a significant process in the atmospheres of Mars and Venus for electron impact on $CO_2$. This process does not appear to have been evaluated for CO, although the density of CO exceeds that of $CO_2$ in the upper atmospheres of these planets. In this paper electron cooling rates for CO are calculated and compared with existing rates for $CO_2$. It is found that electron cooling by CO becomes more significant than by $CO_2$ above altitudes of about 300 km on Mars and about 168 km on Venus. The sensitivity of the calculated cooling rates to different measurements of the integral cross sections for electron-impact vibrational excitation of CO is also investigated.

## Background

Morrison and Greene [1] identified electron cooling by electron impact excitation of $CO_2$ as an important energy transfer process in the atmospheres of Mars and Venus. They presented electron energy loss rates (*i.e.* cooling rates per unit electron and molecule density) as a function of electron temperature. The rate was calculated at each temperature by averaging over the electron energy distribution, assumed to be Maxwellian. These rates can be multiplied by the electron and $CO_2$ densities to give the electron cooling rates in an actual atmosphere. Strangeway [2] used the electron energy loss rates of Morrison and Greene in modeling the ionosphere of Venus.

Measurements have been made of the densities of $CO_2$ and CO in the atmospheres of Mars and Venus. Laboratory measurements have been made of the integral cross sections for vibrational excitation of CO by low-energy electrons. Hence it is now possible to assess the extent of electron cooling by CO, relative to that by $CO_2$. In this paper we review the available data for the cross sections and atmospheric parameters, then use them to calculate electron cooling rates in the atmospheres of both planets. It is found that electron cooling by CO is greater than by $CO_2$ above altitudes of 300 km on Mars and above 168 km on Venus.

The sensitivity of the results to different available cross sections is also investigated. At heights where cooling by CO is significant compared to that of $CO_2$, differences of up to 35% arise from using different cross-section sets for CO. At lower altitudes, where the cooling by CO is much smaller than that of $CO_2$, the differences due to different cross-section sets are much larger. While this is not of any great consequence in the calculation of cooling rates, it is expected to be important in predictions of infrared emissions by CO.

## Results and discussion
### *Electron impact vibrational excitation of CO*
Integral cross sections for electron impact vibrational excitation of CO ($0 \rightarrow v'$) were measured by Schulz [3] for $v' = 1 - 8$ and by Ehrhardt *et al.* [4] for $v = 1 - 7$. Schulz did not give absolute values for individual vibrational levels, but it can be deduced by analysis of the presented graphs that Schulz measured a maximum cross section for $0 \rightarrow 1$ of $3.6 \times 10^{-16}$ cm$^2$ at 1.75 eV. Schulz suggested that this could be in error by a factor of 2. Ehrhardt *et al.* deduced a maximum cross section for $0 \rightarrow 1$ of $3.5 \times 10^{-16}$ cm$^2$ at ~1.83 eV. Subsequent measurements for $0 \rightarrow 1$ excitation were reviewed by Brunger and Buckman [5], including absolute measurements by Gibson *et al.* [6] which give a cross section of $4.874 \times 10^{-16}$ cm$^2$ at 1.91 eV. Recently Poparić *et al.* [7] made measurements for $v' = 1 - 10$, but normalised to the data of Gibson. *et al.* Hence there is only one recent measurement of the absolute integral cross section, which is ~35% higher than earlier values.

Relative measurements for $v' = 1 - 15$ were made by Allan [8]. These are shown in figure 1, scaled to the absolute $0 \rightarrow 1$ value determined by Gibson *et al.* at 1.91 eV. Results for some levels were omitted by Allan, presumably for clarity in the plot, so these have been interpolated here. The measurements of $0 \rightarrow 1$ excitation by Poparić *et al.* [7] are also plotted. The similarity in the absolute values is expected, given that both data sets have been scaled (albeit in different ways) to the data of Gibson *et al.*. The main significant difference in shape is that Allan's cross sections are higher at low energies. Poparić *et al.* observe that their measurements for $v' = 2 - 7$ are higher (~40%) than those of Allan (this being of course subject to the normalisation for $v' = 1$).

### *Atmospheric Parameters*
Atmospheric parameters of Mars and Venus have been measured by various spacecraft since 1976. The measurements required to calculate electron cooling rates are the electron temperature and the densities of $CO_2$, CO and for the free electrons.

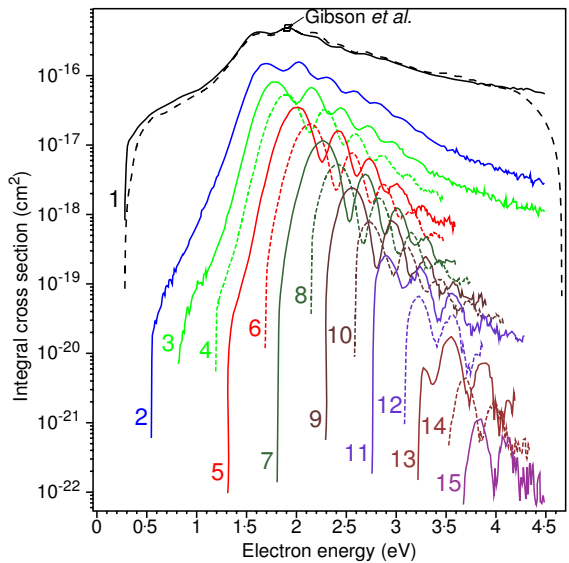

**Figure 1**
**ICS for vibrational excitation of CO**. Integral cross sections for electron-impact vibrational excitation of CO for $0 \to v'$, for $v'$ = 1, 2, 3, 5, 7, 9, 11, 13, 15, determined as relative values by Allan [8] and scaled to the value of Gibson et al [6] for $0 \to 1$ at 1.91 eV (———). Interpolated cross sections for $v'$ = 4, 6, 8, 10, 12, 14 (- - - -). Measurements of Poparić et al. [7] (– – – – –).

In situ measurements of electron temperature and densities of electrons, $CO_2$ and CO in the atmosphere of Mars were made by the Viking 1 lander during its descent in 1976 [9,10]. More recently remote measurements of electron density were made by the Mars Global Surveyor [11]. In Figure 2 the densities of $CO_2$ and CO measured by Viking 1 are plotted, along with fitted functions specified by Nier et al. [9]. These fitted functions were used for the calculations described below. (The regular deviation of the CO measurements from the straight-line fit is not seen in similar measurements by Viking 2 and so may be due to local conditions or uncertainties in the measurements. However, as the Viking 2 electron data cannot be used [10], only Viking 1 measurements of CO density are used here in order to have a consistent set of measurements.) Measurements of the electron density by Viking 1 [10] are shown in Figure 3, along with the maximum and minimum values measured remotely over a 17-day interval by the Mars Global Surveyor [11]. Our fit to these various measurements, labelled "Fit to measurements", was used for the electron densities in the calculations. Measurements of the electron temperature by Viking 1 [10] are shown in figure 4. (These are for thermal electrons only and not for the smaller populations of photoelectrons and high temperature electrons identified by Hanson and Mantas [10].) Due to the absence of data below 200 km, an electron temperature profile from a model by Choi et al. [12], also shown in figure 4 and consistent with the Viking measurements, is used in the calculations for all heights.

Measurements by the "Pioneer Venus" orbiter of the densities of CO2 and CO in the atmosphere of Venus are shown in figure 5, along with densities from an empirical model designated

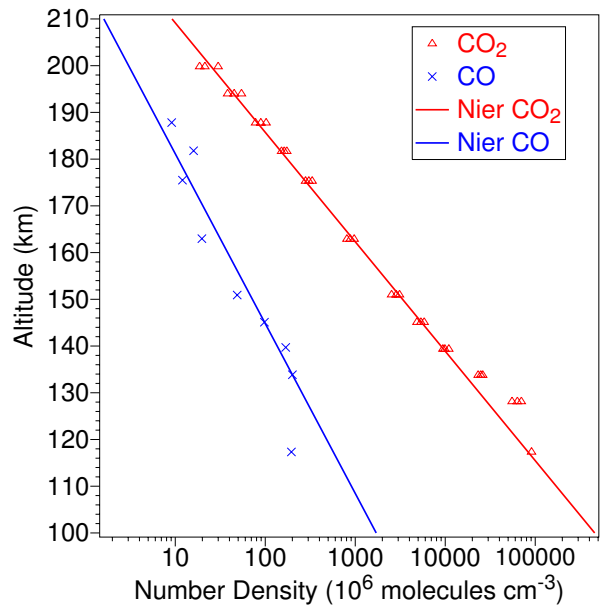

**Figure 2**
**Martian atmospheric densities**. Densities measured by Viking 1 of CO (×) and $CO_2$ (△), and the fits to these of Nier *et al.* [9]. Density is plotted on the horizontal axis against height in the Martian atmosphere on the vertical axis.

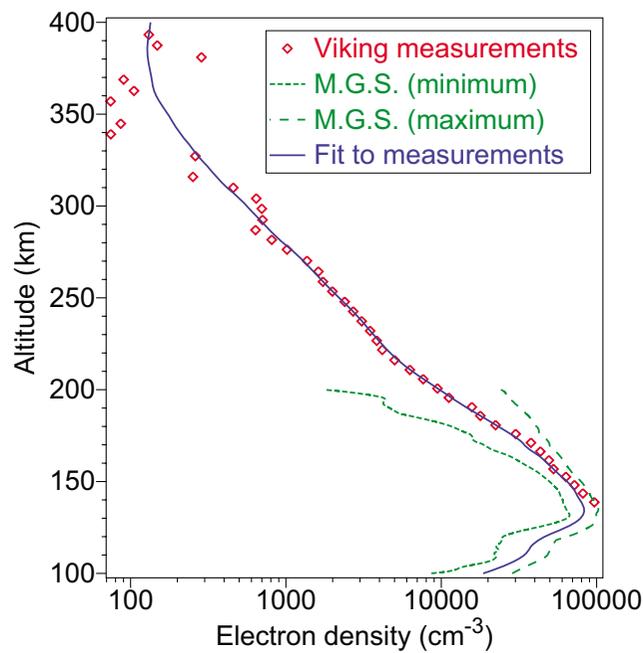

**Figure 3**
**Martian electron densities**. Electron densities measured by Viking 1 [10] (◊), the minimum (- - -) and maximum (– – –) values observed by Mars Global Surveyor [11], and the fit to the measurements (——) used in this work.

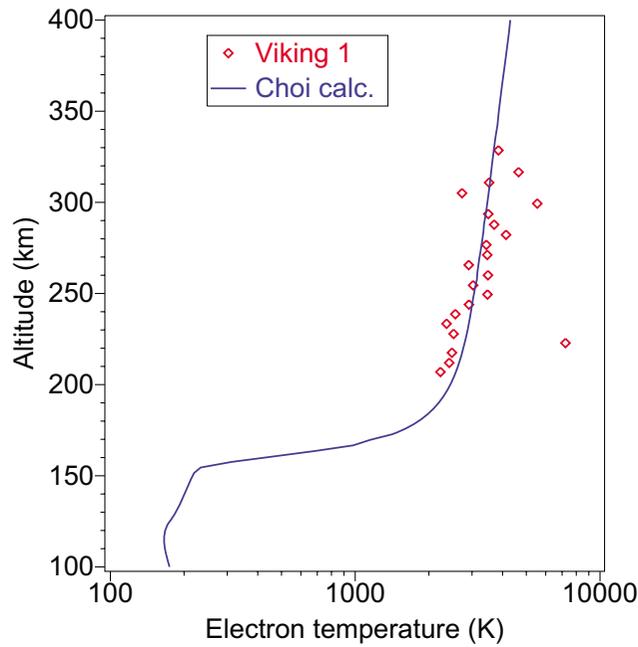

**Figure 4**
**Martian electron temperatures**. Electron temperatures measured by Viking 1 [10] (◊) and the modeled values of Choi *et al.* [12] (——).

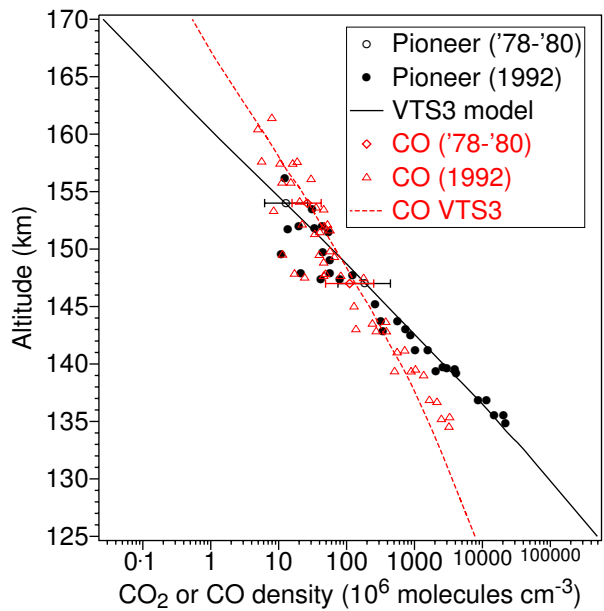

**Figure 5**
**Venusian atmospheric densities**. Densities [13] measured by Pioneer Venus in '78–'80 of $CO_2$ (○) and CO (◊) and in 1992 for $CO_2$ (●) and CO (△), and an empirical model "VTS3" for $CO_2$ (——) and CO (– – –).

"VTS3" [13]. The model densities were used in the calculations. Measurements of the electron densities and temperatures by Pioneer Venus [14] are shown in figures 6 and 7. As there are no measurements below 150 km, calculated values from a model by Strangeway [2] are used for the lower altitudes. Only measurements near local midnight are included, to be consistent with the conditions of Strangeway's model. Curves fitted to the measurements and Strangeway's model are shown in figures 6 and 7 and were used in the calculations.

*Electron energy transfer rates*

Electron energy transfer rates (these being the energy transferred in one second per unit electron and molecule density) for the vibrational excitations of $CO(0 \rightarrow v')$ for $v' = 1 - 13$ were calculated using the formulation of Pavlov [15] and are shown in figure 8. The electron impact cross sections used were the relative values of Allan [8], scaled as shown in figure 1. Also shown are the rates for $CO(0 \rightarrow 1)$ calculated using the cross sections of Poparić *et al*. The latter give smaller energy transfer rates at electron temperatures below ~5000 K. The sum of the energy transfer rates using Allan's cross sections is shown in figure 8 along with the sum for vibrational modes of $CO_2$ calculated by Morrison and Greene [1]. The sum for CO is very much less than for $CO_2$ at lower electron temperatures, but higher above ~5000 K.

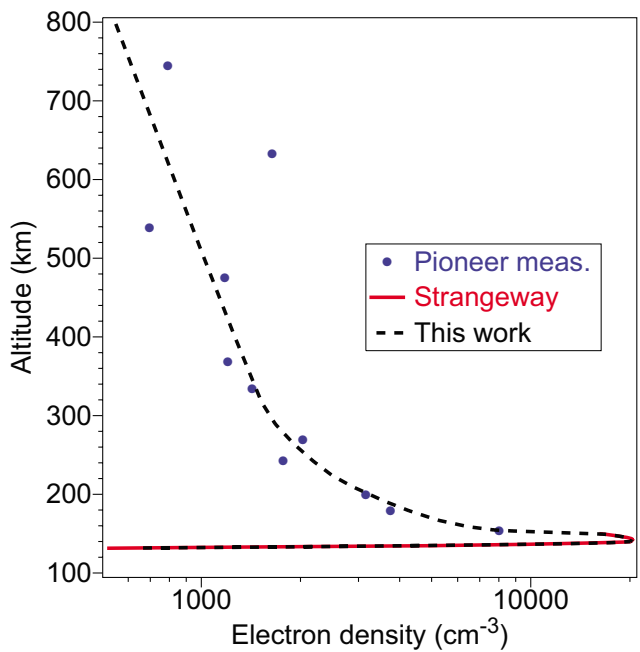

**Figure 6**
**Venusian electron densities**. Electron densities measured by Pioneer Venus [14] (●), calculated in the model of Strangeway [2] (———) and the fit (– – –) used in this work.

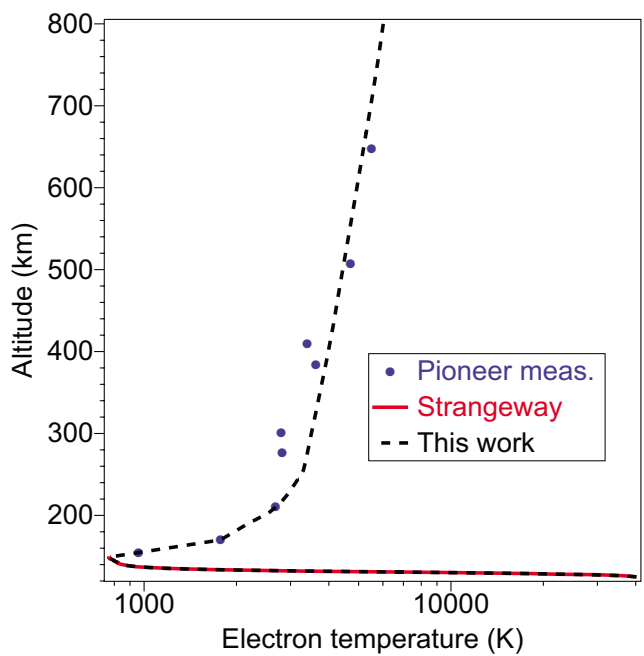

**Figure 7**
**Venusian electron temperatures**. Electron temperatures measured by Pioneer Venus [14] (●), calculated in the model of Strangeway [2] (——) and the fit (– – –) used in this work.

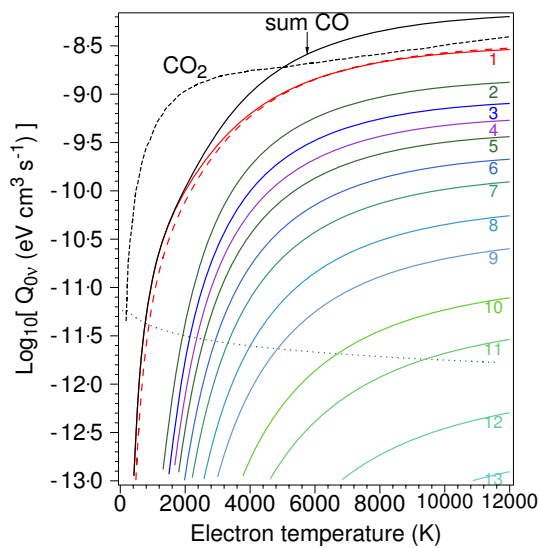

**Figure 8**
**Electron energy transfer rates**. $\text{Log}_{10}$ of electron energy transfer rates as a function of electron temperature for electron impact excitation of $CO(0 \rightarrow v)$ for $v = 1 - 13$ (—— as labelled), for their sum ("sum CO") and for $CO_2$ (- - -) as given by Morrison and Greene [1]. The rate for $CO(0 \rightarrow 1)$ calculated with the cross sections of Poparić *et al.* [7] is also shown (– – –), as is the rotational excitation ($J = 0 \rightarrow 1$) using the cross sections of Randell *et al.* [16] (· · ·).

The transfer rates due to rotational excitation ($J = 0 \to 1$), calculated using the cross sections presented by Randell *et al.* [16], are also shown in figure 8. These rates are small compared to the $0 \to 1$ vibrational excitations, except at very low electron temperatures.

### *Calculation of cooling rates*

Electron cooling rates (*i.e.* the electron energy transfer rates multiplied by the electron and gas densities) are plotted in figure 9 for the atmosphere of Mars. The rates are shown for $CO_2$ (calculated using the rates of Morrison and Greene [1]) and for CO using the cross sections of Allan [8] (scaled as described earlier) and those of Poparić *et al.* [7]. The rates for CO are shown separately for the $0 \to 1$ excitation and for the sum of rates for $0 \to v'$, where $v' = 2 - 10$. The apparent oscillatory structures for $CO_2$ and CO are due to the interplay of the various changes in atmospheric parameters, such as the rapid decrease in electron temperature with height at ~165 km and the peak in the electron density at ~135 km.

The cooling rates due to the rotational excitation $J = 0 \to 1$ in CO are also shown in figure 9. They are not significant relative to those of $CO_2$ at any height and so are not considered further.

The cooling rates due to $CO(0 \to 1)$ are substantially less than those for $CO_2$ below ~280 km, but are greater above 300 km. The cooling rates due to $CO(0 \to v', v' = 2 - 10)$ are much smaller again at the lower altitudes but exceed those of $CO_2$ at altitudes above ~310 km. At the heights where the CO cooling is significant, there is little difference between the rates for $0 \to 1$ excitation for the two different cross-section sets, as expected given their similarity at higher temperatures. For excitation to higher levels the cooling rates are about 35% higher for the cross sections of Poparić *et al.*, consistent with their higher values for the cross sections. As both of these comparisons are dependent on the same normalisation to the values of Gibson *et al*, it would be useful if another independent measurement of the absolute cross sections for CO could be made.

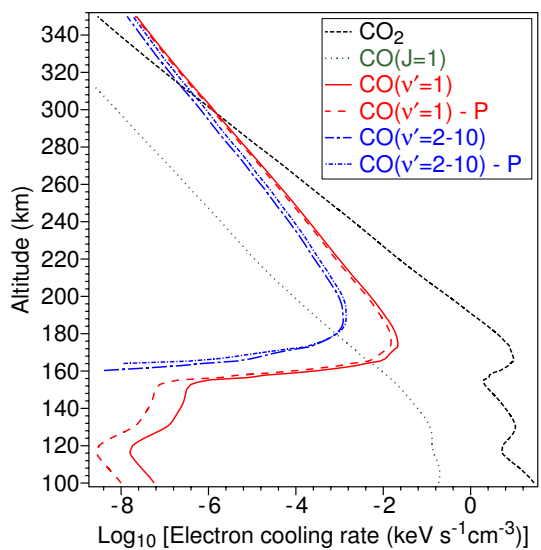

**Figure 9**
**Electron Cooling rates for $CO_2$ and CO at Mars**. $Log_{10}$ of electron cooling rates (horizontal axis), as a function of altitude in the atmosphere of Mars, for $CO_2$ using the transfer rates of Morrison and Greene [1] (- - -), for rotational ($J = 0 \to 1$) excitation of CO (· · ·), for $CO(0 \to 1)$ using the cross sections of Allan [8] (——) and Poparić *et al.* [7] (– – –), and for $CO(0 \to v', v' = 2 - 10)$ using the cross sections of Allan (—— –) and Poparić *et al.* (- · - · -).

The cross sections of Allan lead to higher cooling rates in the altitude range 100–156 km for the $0 \to 1$ excitation and below ~168 km for the higher levels ($v' = 2 - 10$). As at these heights the CO cooling rates are insignificant compared to those of $CO_2$, these differences are inconsequential as far as calculations of cooling rates are concerned. However, as the infrared emissions due to radiative decay following excitation are related to the excitation rate, the differences are worth noting in case the infrared emissions from CO can be measured with sufficient sensitivity. In this case it would be useful to resolve the differences at low energy between the measurements of Allan and those of Poparić *et al*.

The same set of results is shown for the atmosphere of Venus in figure 10. Again the CO contribution to the cooling rate is much smaller than for $CO_2$ in the lower altitude range, but is greater at heights above ~168 km for $0 \to 1$ excitation and above ~184 km for $0 \to v'$, $v' = 2 - 10$. There are differences of up to ~20% in using the cross sections of Allan or Poparić *et al.* at heights where the cooling rate due to CO is significant. However, there is a much larger difference (up to a factor of more than 10) for the two sets of cross sections for the higher vibrational excitations ($v' = 2 - 10$) in the altitude range 138–158 km, corresponding to electron temperatures of ~700–1000 K. Thus, as for Mars, it would be useful to verify both the absolute value of the CO cross sections for calculations of electron cooling at higher altitudes and the low-energy cross-section shape for modeling of CO emissions at lower altitudes.

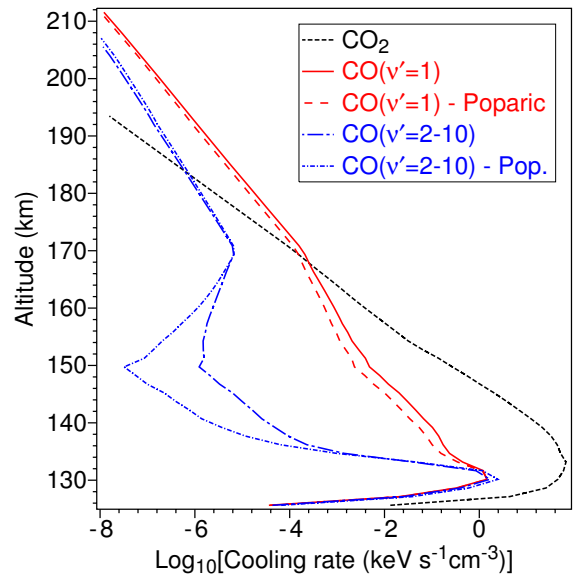

**Figure 10**
**Electron Cooling rates for $CO_2$ and CO at Venus**. $Log_{10}$ of electron cooling rates (horizontal axis), as a function of altitude in the atmosphere of Venus, for $CO_2$ using the transfer rates of Morrison and Greene [1] (- - -), for $CO(0 \to 1)$ using the cross sections of Allan [8] (———) and Poparić *et al.* [7] (– – –), and for $CO(0 \to v'$, $v' = 2 - 10)$ using the cross sections of Allan (——— –) and Poparić *et al.* (- · - · -).

## Conclusion

Electron cooling rates due to CO in the atmospheres of Mars and Venus have been calculated using two sets of electron impact cross sections for vibrational excitation. It was found that above ~300 km in the atmosphere of Mars and above ~168 km in the atmosphere of Venus, the cooling rates due to $CO(0 \rightarrow 1)$ exceed the total rate for all excitations of $CO_2$. In each case at a slightly greater height the cooling rate due to electron-impact excitation to the higher vibrational levels of CO is also greater than for all levels of $CO_2$. At the heights where the cooling rate due to CO is significant, there are differences of up to ~35% between the rates calculated with the two different cross-section sets, so more definitive values of the absolute CO cross sections would be useful for accurate modeling of cooling rates. There are much larger differences between the cooling rates calculated with the different cross section sets at altitudes with lower electron temperatures. While these differences are of no concern for the calculation of electron cooling rates, which are dominated by $CO_2$ at these altitudes, they would make a difference to the calculated CO infrared emissions. In this case it would be useful to resolve the differences between experimental measurements of the CO cross sections at low energy.

## Acknowledgements
This work was supported by the Australian Research Council.